\documentclass[twocolumn,trackchanges]{aastex62}
\usepackage{natbib}
\usepackage{empheq}
\usepackage{hyperref}
\usepackage{epstopdf}
\usepackage{amsmath,amsfonts,amssymb}
\usepackage{graphicx}
\begin{document}

\title{The Equation of State and Some Key Parameters of Neutron Stars: Constraints from GW170817, the Nuclear Data, and the Low-mass X-ray Binary Data}
\author{Jin-Liang Jiang}
\author{Shao-Peng Tang}
\author{Dong-Sheng Shao}
\author{Ming-Zhe Han}
\author{Yin-Jie Li}
\affil{Key Laboratory of dark Matter and Space Astronomy, Purple Mountain Observatory, Chinese Academy of Sciences, Nanjing, 210023, China.}
\affil{School of Astronomy and Space Science, University of Science and Technology of China, Hefei, Anhui 230026, China.}
\author{Yuan-Zhu Wang}
\affil{Key Laboratory of dark Matter and Space Astronomy, Purple Mountain Observatory, Chinese Academy of Sciences, Nanjing, 210023, China.}
\author{Zhi-Ping Jin}
\author{Yi-Zhong Fan}
\author{Da-Ming Wei}
\affil{Key Laboratory of dark Matter and Space Astronomy, Purple Mountain Observatory, Chinese Academy of Sciences, Nanjing, 210023, China.}
\affil{School of Astronomy and Space Science, University of Science and Technology of China, Hefei, Anhui 230026, China.}
\email{tangsp@pmo.ac.cn (SPT) and yzfan@pmo.ac.cn (YZF)}

\begin{abstract}
In this work we parameterize the equation of state of dense neutron star (NS) matter with four pressure parameters of $\{\hat{p}_1, \hat{p}_2, \hat{p}_3, \hat{p}_4\}$ and then set the combined constraints with the data of GW 170817 and the data of six low-mass X-ray binaries (LMXBs) with thermonuclear burst or alternatively the symmetry energy of the nuclear interaction. We find that the nuclear data effectively narrow down the possible range of $\hat{p}_1$, the gravitational-wave data plays the leading role in bounding $\hat{p}_2$, and the LMXB data as well as the lower bound on the maximal gravitational mass of non-rotating NSs govern the constraints on $\hat{p}_3$ and $\hat{p}_4$. Using posterior samples of pressure parameters and some universal relations, we further investigate how the current data sets can advance our understanding of tidal deformability ($\Lambda$), moment of inertia ($I$), and binding energy (BE) of NSs. For a canonical mass of $1.4M_\odot$, we have $I_{1.4} = {1.43}^{+0.30}_{-0.13} \times 10^{38}~{\rm kg \cdot m^2}$, $\Lambda_{1.4} = 390_{-210}^{+280}$ , $R_{1.4} = 11.8_{-0.7}^{+1.2}~{\rm km}$ and $BE_{1.4} = {0.16}^{+0.01}_{-0.02} M_{\odot}$  if the constraints from the nuclear data and the gravitational-wave data have been jointly applied. For the joint analysis of gravitational-wave data and the LMXB data, we have $I_{1.4} = {1.28}^{+0.15}_{-0.08} \times 10^{38}~{\rm kg \cdot m^2}$, $\Lambda_{1.4} = 220_{-90}^{+90}$, $R_{1.4} = 11.1_{-0.6}^{+0.7}~{\rm km}$, and $BE_{1.4} = {0.18}^{+0.01}_{-0.01} M_{\odot}$. These results suggest that the current constraints on $\Lambda$ and $R$ still suffer from significant systematic uncertainties, while $I_{1.4}$ and $BE_{1.4}$ are better constrained.
\end{abstract}

\section{Introduction}

As the compact objects contain material with the highest densities in the observable universe, neutron stars (NSs) serve as the ideal laboratories for studying extremely dense matter \citep[see, e.g.][for recent reviews]{2012ARNPS..62..485L,2016PhR...621..127L,2016ARA&A..54..401O,2017RvMP...89a5007O}. So far, about $2000$ NSs, mainly consisting of  pulsars\footnote{http://www.atnf.csiro.au/research/pulsar/psrcat}, have been measured in the Galaxy. The measurements of masses and/or radii for a small fraction of NSs have set interesting constraints on the properties of the very dense matter. For example, the detections of a few NSs with a gravitational mass of $\approx 2M_\odot$ \citep{ 2010Natur.467.1081D, 2013Sci...340..448A, 2019NatAs.tmp..439C} have excluded the soft equations of states (EoSs) that are unable to support such massive objects. For some NSs in the LMXB systems, there is a good opportunity to measure their radius and mass simultaneously via spectroscopic observation of the thermonuclear burst that happened on their surfaces, or through observations of their angular size when they remain in the quiescent state \citep[see, e.g.][for a comprehensive review]{2016ARA&A..54..401O}. These radii/masses data have been widely adopted to constrain the EoS of ultra-high-dense matter \citep{2009PhRvD..80j3003O, 2010ApJ...722...33S, 2013ApJ...765L...5S, 2014EPJA...50...40L, 2014ApJ...784..123L, 2016A&A...591A..25N, 2016ApJ...820...28O, 2017ApJ...844..156R, 2019arXiv190501081B, 2019arXiv190205078F}.

Nuclear experiments are also progressively narrowing down the ranges of parameters that describe the symmetry energy near the nuclear saturation density, which can be further adopted to infer the physical properties of NSs \citep{2013ApJ...771...51L, 2018PhRvL.121f2701L, 2019JPhG...46g4001K} including, for instance, the radii, moments of inertia, and the binding energy.

The discovery of the first NS merger-driven gravitational-wave event GW170817 \citep{2017PhRvL.119p1101A} has provided the community with a valuable/novel opportunity to reliably probe the EoS and the NS properties. In particular, with some reasonable assumptions and EoS-independent relationships, the tidal deformabilities and the radii of the two NSs involved in GW170817 have been measured and some bulk properties of NSs have been inferred \citep[e.g.,][]{2018PhRvL.120q2703A, 2018PhRvL.120q2702F, 2018PhRvL.120z1103M, 2018PhRvL.121f2701L,2018PhRvL.121i1102D, 2018PhRvL.121p1101A, 2018ApJ...868L..22L, 2019arXiv190205502L, 2019PhRvD..99l3026K}.

Inspired by the above remarkable advances, in this work we try to further explore the potential of constraining the EoS of dense NS matter with the mass ($M$) and/or radius ($R$) measurements of the NSs, the nuclear experimental data, and GW170817. Special attention is paid to the dependence of the results on the data set adopted in the investigation.

This work is organized as follows. In Section \ref{sec:methods} we introduce the methods. The results on the EoS constraints and bulk properties of NSs are presented in Section \ref{sec:results}. Section \ref{sec:sum_cons} is our summary and discussion.


\section{Methods}
\label{sec:methods}

\subsection{Parameterizing EoS}
\label{sec:para-eos}

Parameterized representations of the EoS play a very important role in efforts to measure the properties of the matter in the cores of NSs using astronomical observations and the gravitational-wave data. A number of methods to effectively parameterize the realistic EOS models have been developed in the literature \citep{2010PhRvD..82j3011L, 2014ApJ...789..127K, 2016EPJA...52...18S, 2019arXiv190205502L, 2019arXiv190404233M}, including, for instance, the spectral expansion \citep{2010PhRvD..82j3011L} and the piecewise polytropic expansion \citep{2009PhRvD..79l4032R, 2016ARA&A..54..401O, 2017ApJ...844..156R}.

Usually, the piecewise polytropic expansion can be carried out in four ways. The first approach is to introduce a set of pressures at given densities to approximate the EoS \citep{2016ARA&A..54..401O, 2017ApJ...844..156R}. The second is to adopt a series of adiabatic indexes in given density ranges \citep{2009PhRvD..79l4032R}. The third is to parameterize pressure difference between two neighboring fixed densities \citep{2016EPJA...52...18S}. The last is to parameterize densities and pressures
simultaneously \citep{2016EPJA...52...18S}. In each case, the EoS in each density range can be expressed as
\begin{equation}
  \label{eq:piece_eos}
  P = K\rho^{\Gamma},
\end{equation}
where $P$ is the pressure, $\rho$ is the mass density, $K$ is constant in each piece of EoS, and $\Gamma$ is the adiabatic index. Here, we adopt the first method by parameterizing EoS using four pressures $\{P_1, P_2, P_3, P_4\}$ at the corresponding densities of $\{1, 1.85, 3.7, 7.4\}\rho_{\rm sat}$ \citep{2009PhRvD..80j3003O}, where $\rho_{\rm sat}=2.7 \times 10^{14} \rm g/cm^3$ is the so-called saturation density.

With a specific parameterized EoS in hand, we need one additional parameter, the central pseudo-enthalpy ($h_{\rm c}$), to determine the global properties of non-rotating NSs such as  the gravitational mass $M$, the mean radius $R$ and the dimensionless tidal deformability $\Lambda$, etc. The $h_{\rm c}$ is defined as
\begin{equation}
  \label{eq:hc}
  h_{\rm c} \equiv \int_{0}^{p_{\rm c}} \frac{dp}{\epsilon(p)+p},
\end{equation}
where $p$ is the pressure, $\epsilon$ is the energy density, and $p_{\rm c}$ is the pressure at the center of the NS.

We implement method described in Appendix C of \citet{2014PhRvD..89f4003L} to calculate the global properties $\{M,R,\Lambda\}$ from parameters $\{h_{\rm c}, P_1, P_2, P_3, P_4\}$. A common EoS table for $\rho \leq 0.33\rho_{\rm sat}$ is adopted from SLy EoS table \citep[][]{2016ApJ...820...28O}\footnote{\label{xtreme}http://xtreme.as.arizona.edu/NeutronStars}.

For convenience, hereafter we replace the EoS parameters $\{P_1, P_2, P_3, P_4\}$ with the equivalent dimensionless parameters $\{\hat{p}_1, \hat{p}_2, \hat{p}_3, \hat{p}_4\}$, where $\hat{p}_i =  P_i / (10^{32+i}\,{\rm dyn~cm^{-2}})$.

\subsection{Priors and known constraints of the EoS parameters}
\label{sec:nuc-cons}

We use a flat prior for every pressure parameters unless with a specific statement. The ranges of these parameters are set to be consistent with realistic EoSs of dense matter shown in \citet{2009PhRvD..79l4032R} \footnote{The exception is that for Test F we have to take significantly wider prior distributions of the pressure parameters, otherwise it is not possible to well reproduce the $M$-$R$ distributions reported in \citet{2016ApJ...820...28O}, as very small radii were suggested for a few sources.
Most of the ``enlarged" regions, however, are found to be rejected by physical conditions such as causality and $M_{\rm TOV}$ limit.
}, namely $\hat{p}_1 \in [1.5, 13.5]$, $ \hat{p}_2 \in [0.7, 8.0]$, $\hat{p}_3 \in [0.6, 7.0]$, and $\hat{p}_4 \in [0.3, 4.0]$. In additionally, the EoS parameters $\{P_1, P_2, P_3, P_4\}$ should satisfy the following constraints \citep{2017ApJ...844..156R}:

(i) The microscopical stability, i.e., $P_4 \ge P_3 \ge P_2 \ge P_1$.

(ii) The physically plausible condition of causality, i.e.,
\begin{equation}
    \label{causal}
    \frac{c^{2}_{\rm s}}{c^2} = \frac{dp(h)}{d\epsilon(h)}\le 1 ~~~{\rm for}~~~h\le h_{\rm c,max},
\end{equation}
where $p(h)$, $\epsilon(h)$, $h$, and $c$ are pressure, energy density, pseudo-enthalpy, speed of the light, respectively. $h_{\rm c,max}$ is the central enthalpy of a non-rotating stable NS with a maximal gravitational mass ($M_{\rm TOV}$).

(iii) Maximum stable mass of non-rotating NS ($M_{\rm TOV}$) is likely within the range of $[2.06, 2.5]M_{\odot}$ \citep{1998PhRvC..58.1804A, 2016PhR...621..127L} \footnote{In the literature, some tighter bounds on $M_{\rm TOV}$ have been suggested \citep[e.g.,][]{Fan2013}. However, these bounds are highly model-dependent and in the current analysis we do not take them into account.}. The lower limit is taken to be slightly smaller than the $68.3\%$ lower limit of the mass of \objectname{PSR J0740+6620} \citep{2019NatAs.tmp..439C}, i.e., $2.07 M_{\odot}$. This is because \objectname{PSR J0740+6620} has a rotation frequency of $346.532$ Hz, which can slightly weaken the constraints on the $M_{\rm TOV}$ to a value of $2.06M_{\odot}$ \citep[see][for some relevant discussions]{2016MNRAS.459..646B, 2018ApJ...858...74M}.

(iv) The adiabatic indexes in all the plausible density regions should satisfy the condition $\Gamma<7$ \citep{2019arXiv190205078F}.

\subsection{LMXB data}

As mentioned in Section \ref{sec:para-eos}, given a set of parameters $\{h_{\rm c}, \hat{p}_1, \hat{p}_2, \hat{p}_3, \hat{p}_4\}$, one can derive the mass and the radius of an NS. While the observations of an LMXB system yield the probability distribution function of the masses and the radii of the NSs. Thus, if we take these sources into consideration, the likelihood for these galactic NSs then takes the form
\begin{equation}
    \label{eq:like_gn}
    L_{GN}({\vec{\theta}_{GN}}) = \prod_{i=1}^{n} P_{\rm i}(M(\vec{\theta_{\rm i}}), R(\vec{\theta_{\rm i}})),
\end{equation}
where $n$ is the number of NSs taken in this analysis, $P_i$ is the likelihood at $\{M(\vec{\theta_i}), R(\vec{\theta_i})\}$ interpolated from the likelihood table of the $i$th source \citet{2016ApJ...820...28O}, and $\vec{\theta_{\rm i}} := \{h_{\rm ci}, \hat{p}_1, \hat{p}_2, \hat{p}_3, \hat{p}_4\}$ are the basic parameters to describe a cold non-rotating NS. Thus, the ${\vec{\theta}_{GN}}$ can take the form \[\begin{split}
    \vec{\theta}_{GN} := \cup_{i=1}^n \vec{\theta_{\rm i}} &= \{h_{\rm ci}| i= 1,2,3,...,n\} \\
    &\cup \{\hat{p}_1, \hat{p}_2, \hat{p}_3, \hat{p}_4\},
\end{split}\]
which contains two parts, i.e., four EoS parameters and the central enthalpies of $n$ NSs.

Six sources, namely \objectname{4U 1820-30}, \objectname{4U 1724-207}, \objectname{EXO 1745-248}, \objectname{SAX J1748.9-2021}, \objectname{KS1731-260} and \objectname{4U 1608-52}, whose masses and radii are constrained by thermonuclear burst data, have been taken into account in our analysis. The masses, radii, and associated likelihood data are directly taken from \citet[][see the web source (see footnote 2)]{2016ApJ...820...28O}. The pseudo-enthalpy at the center of each galactic NS $h_{\rm ci}$ is assumed to span uniformly in the range $[0.1, 0.8]$.

\subsection{Symmetry energy}
\label{sec:sym_energy}

We know that nuclear experiments can also contribute to constraining the EoS parameters \cite[e.g.][]{2013ApJ...771...51L}. In $\beta$-equilibrium condition, the pressure of matter at neutron saturation density satisfies \citep{2014EPJA...50...40L}
\begin{equation}
    \label{pressure_at_saturation}
    p_{\beta}(n_s)\simeq\frac{L}{3}n_s\left[1-{\left(\frac{4S_v}{\hbar c}\right)}^3\frac{4-3S_v/L}{3 \pi^2 n_s}+\ldots\right]
\end{equation}
where $n_s=0.16\ {\rm fm}^{-3}$ is the saturation baryon number density, $(S_v, L)$ are symmetry parameters, and $\hbar$ is the reduced Planck constant. We take the bounds on $S_v, L$ found in \citet{2017ApJ...848..105T}, and apply a similar process used in \citet{2014EPJA...50...40L} to transform these constraints to a distribution of $p_{\beta}(n_s)$ using equation (\ref{pressure_at_saturation}) with Monte Carlo sampling. Then, we can obtain a $95\%$ confidence interval of pressure $\hat{p}_1$ at $n_s$, which is $[3.12,~4.70]$. Meanwhile, \citet{2016ApJ...820...28O} also found a constraint $P_2>7.56\,{\rm MeV\,~ fm^{-3}}$ using nuclear data, which corresponds to our parameter $\hat{p}_2>1.21$. If we take these nuclear constraints into analysis, the likelihood for these constraints should read
\begin{equation}
    \label{eq:like_nuc}
    L_{Nuc}(\vec{\theta}_{Nuc}) = \begin{cases}
                   &1~~~~(3.12 < \hat{p}_1 < 4.70,~\hat{p}_2 > 1.21) \\
                   &-\infty~~~~{\rm (Otherwize)}
              \end{cases},
\end{equation}
where $\vec{\theta}_{Nuc} := \{\hat{p}_1, \hat{p}_2, \hat{p}_3, \hat{p}_4\}$.

\subsection{GW data}
\label{sec:gw}

Instead of sampling the parameters $\{M_1, M_2, \Lambda_1, \Lambda_2\}$ in usual analyses of GW170817 data, we sample $\{\mathcal{M}_{\rm c}, q, \hat{p}_1, \hat{p}_2$, $ \hat{p}_3, \hat{p}_4\}$ to determine the former four parameters. $M_1$ and $M_2$ are determined by equations
\begin{equation}\begin{split}
    M_1 &= q^{2/5}(q+1)^{1/5}\mathcal{M}_{\rm c}, \\
    M_2 &= q^{-3/5}(q+1)^{1/5}\mathcal{M}_{\rm c},
\end{split}\end{equation}
where $\mathcal{M}_{\rm c}$ is chirp mass and $q$ is mass ratio. Note that here $M_1$ and $M_2$ are detector frame parameters, but we can just calculate the source frame masses through parameterized EoSs, so we optimize central enthalpy $h_{\rm c1}$ and $h_{\rm c2}$ to get $M_1/(1+z)$ and $M_2/(1+z)$, respectively, where $z = 0.0099$ is the geocentric redshift of the source of GW170817 \citep{2017ApJ...848L..28L, 2019PhRvX...9a1001A} inferred from the electromagnetic observations of the host galaxy NGC4993. Then we can combine four pressure parameters and optimized central enthalpy $h^{\rm opt}_{\rm c1}$($h^{\rm opt}_{\rm c2}$) to calculate $\Lambda_1$($\Lambda_2$).

We restrict the sky location to the known position of SSS17a/AT 2017gfo \citep{2017ApJ...848L..12A} following \citet{2019PhRvX...9a1001A}, and assume the spin of each NS is aligned with the orbital angular momentum. Additionally, we marginalize phase and distance over likelihood, because they have little correlation with parameters we care about \citep{2012PhRvD..85l2006A, 2019PhRvX...9a1001A, 2019EPJA...55...50R}, and by this mean we can save much time in Markov chain Monte Carlo (MCMC) sampling.

Based on the above considerations, if we take the GW170817 into analysis, the likelihood for the gravitational-wave \citep{2012PhRvD..85l2006A} of each detector would have the functional form
\begin{equation}
    \label{eq:like_gw}
    L_{GW}(\vec{\theta}_{GW}) \propto \exp{[-2\int_0^{\infty} \frac{|\tilde{d}(f)-\tilde{h}(f;\vec{\theta}_{GW})|^2}{S_n(f)}\, df]},
\end{equation}
where
\[ \vec{\theta}_{GW} :=  \{\hat{p}_1, \hat{p}_2, \hat{p}_3, \hat{p}_4\} \cup \{\mathcal{M}_{\rm c}, q, \chi_1, \chi_2, \theta_{\rm jn}, t_{\rm c}, \Psi\}, \]
$\chi_i(i=1, 2)$, $\theta_{\rm jn}$, $t_{\rm c}$ and $\Psi$ are spin magnitudes of NS, the angle between the line of sight and the binary NS system's total angular momentum, the GPS time when the coalescence signal reaches the geocenter of the Earth, and polarization, respectively. $\tilde{d}(f)$, $S_n(f)$, and $\tilde{h}(f;\vec{\theta}_{GW})$ are the Fourier transform of the time domain signal of GW170817, the power spectral density of the data, and the frequency domain strain data generated using the parameter $\vec{\theta}_{GW}$, respectively. And we evaluate this part of the likelihood using PyCBC Inference package \citep{2018ascl.soft05030T, 2019PASP..131b4503B}.

We take publicly available cleaned $4096$ Hz gravitational-wave data\footnote{https://www.gw-openscience.org} lying in the GPS time segment $[1,187,008,682,$ $1,187,008,890]s$ into analysis. An aligned spin prior and a cosine uniform prior of orbital inclination angle $\theta_{jn}$ are adopted. Moreover, $\mathcal{M}_c$, $q$ (the mass ratio), $\chi_i(i=1, 2)$, $\Psi$ and $t_c$ distribute uniformly in the range $[1.18, 1.21]M_{\odot}$, $[0.5, 1.0]$, $[0, 0.05]$, $[0, 2\pi]$ and $[1,187,008,882, 1,187,008,883]$ s, respectively.

\subsection{Joint analysis}

To sample a group of parameters in the MCMC procedure, we need a likelihood describing how probable the data could be given a specific group of parameters and a prior probability of these parameters. Below we examine the roles of different sets of data and/or constraints/assumptions in reconstructing the EoS (see Table \ref{tb:expirement}) by setting different likelihoods/priors in each analysis as follows:

(A) The gravitational-wave data of GW170817 and the default constraints on $M_{\rm TOV}$ (i.e., $\in [2.06, 2.5]M_{\odot}$). The priors of four pressure parameters have been introduced in Section \ref{sec:nuc-cons}. The waveform model PhenomDNRT is adopted. The total likelihood is $L_{GW}(\vec{\theta}_{GW})$ and the number of free parameters is $11$.

(B) The same as scenario (A) except a more ``conservative'' bound on $M_{\rm TOV}$ (i.e., $M_{\rm TOV} \in [1.97, 2.8] M_{\odot}$) is considered.

(C) The same as scenario (A) except that the waveform model TaylorF2 is used.

(D) The same as scenario (A) except that the log-uniform prior for every pressure parameters has been assumed.

(E) In comparison to scenario (A), additional constraints from the symmetry energies have been imposed  (see Section \ref{sec:sym_energy}). The total likelihood is given by
\begin{equation}
    \label{eq:like_gw_nuc}
    L(\vec{\theta}) = L_{GN}(\vec{\theta}_{GW}) \times L_{Nuc}(\vec{\theta}_{Nuc}),
\end{equation}
where
\[\begin{split}
    \vec{\theta} = \vec{\theta}_{GW} \cup \vec{\theta}_{Nuc} = \{\hat{p}_1, \hat{p}_2, \hat{p}_3, \hat{p}_4\} \\
    \cup \{\mathcal{M}_c, q, \chi_1, \chi_2, \theta_{jn}, t_c, \Psi\},
\end{split}\]
 and there are $11$ parameters total.

(F) The same as scenario (A) except for the inclusion of six LMXB sources with thermonuclear bursts. The total likelihood is given by
\begin{equation}
    \label{eq:like_gw_gn}
    L(\vec{\theta}) = L_{GN}(\vec{\theta}_{GN}) \times L_{GW}(\vec{\theta}_{GW}),
\end{equation}
where
\[\begin{split}
    \label{eq:theta_gw_gn}
    \vec{\theta} = \vec{\theta}_{GN} \cup \vec{\theta}_{GW} = \{h_{\rm ci}| i=1,2,3,...,n\} \\
    \cup \{\hat{p}_1, \hat{p}_2, \hat{p}_3, \hat{p}_4\} \cup \{\mathcal{M}_c, q, \chi_1, \chi_2, \theta_{jn}, t_c, \Psi\}.
\end{split}\]
There are $17$ free parameters in total, as we adopt six LMXB sources ($n=6$) in the analysis.

Our MCMC sampling is carried out using the Bilby \citep{2019ascl.soft01011A} built-in sampler PyMultiNest \citep{2016ascl.soft06005B}.

\begin{table*}
  \begin{ruledtabular}
  \caption{Different Tests Designed to Show How Different Assumptions can Affect the EoS Constraint}
  \center\begin{tabular}{lccccccc}
  Condition / Test                & Test A            & Test B          & Test C            & Test D            & Test E            & Test F        \\
  \hline
  LMXB data considered            & No                & No              & No                & No                & No                & Yes           \\
  GW data considered              & Yes               & Yes             & Yes               & Yes               & Yes               & Yes           \\
  Nuclear constraints             & No                & No              & No                & No                & Yes               & No            \\
  Waveform model                  & IMRPDNRT\tablenotemark{a}& IMRPDNRT & TaylorF2          & IMRPDNRT          & IMRPDNRT          & IMRPDNRT      \\
  $M_{\rm TOV}/M_{\odot}$ range       &  $[2.06, 2.5]$    & $[1.97, 2.8]$   & $[2.06, 2.5]$     & $[2.06, 2.5]$     & $[2.06, 2.5]$     & $[2.06, 2.5]$ \\
  Prior of pressures              & Uniform           & Uniform         & Uniform           & Log Uniform       & Uniform           & Uniform       \\
  \end{tabular}
  \tablenotetext{a}{Short note of PhenomDNRT waveform model}
  \label{tb:expirement}
  \end{ruledtabular}
\end{table*}

\begin{figure*}
  \centering
  \includegraphics[width=1\textwidth]{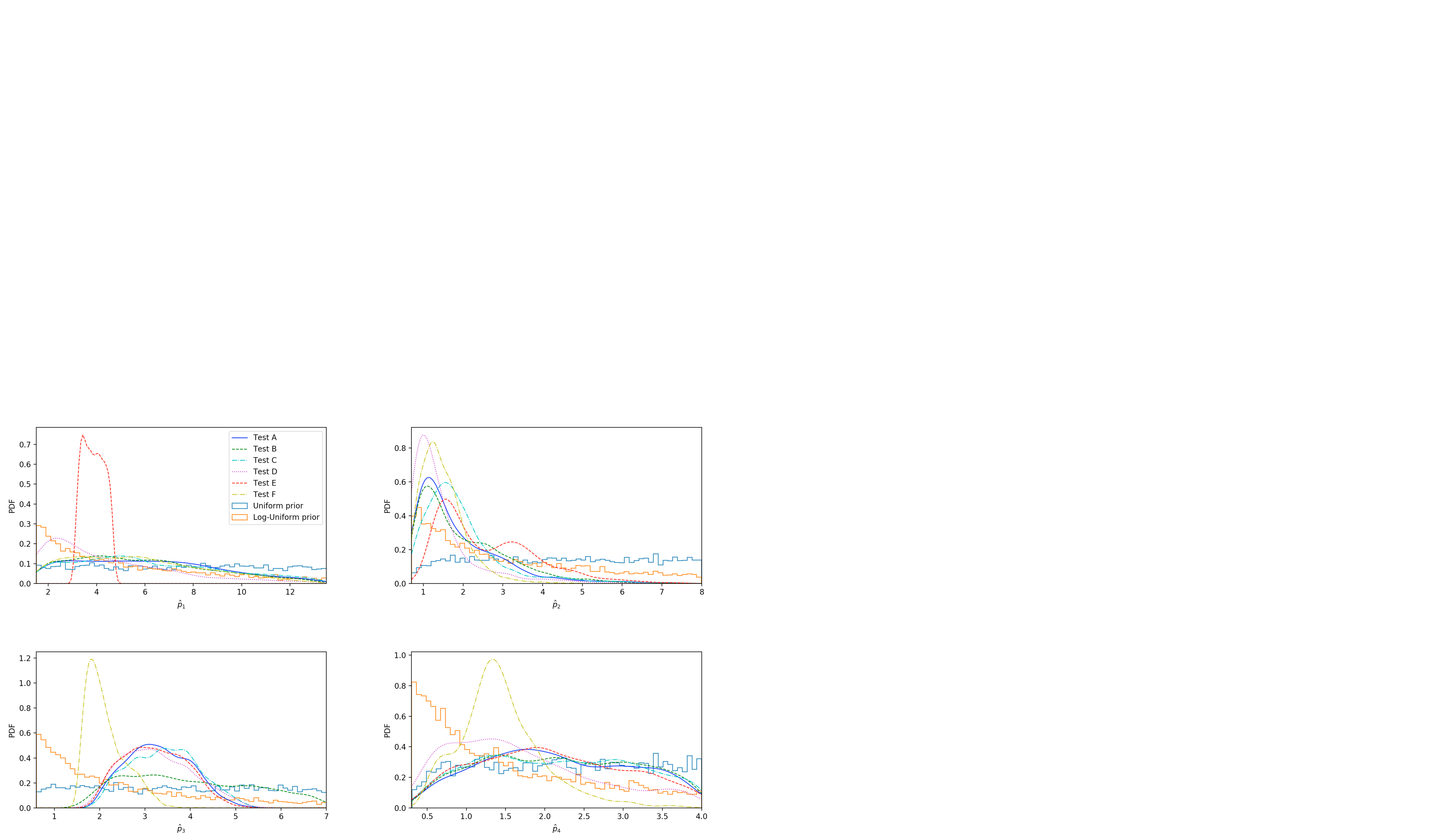}
  \caption{Posterior distributions of four pressure parameters in different tests.}
  \label{fig:post_pressure}
\end{figure*}

\begin{table}
  \begin{ruledtabular}
  \caption{KL-divergence(in bits) Between the Prior and Posterior Distribution for Pressure Parameters in Different Tests}
  \begin{tabular*}{0.3\textwidth}{lcccc}
  Test / $D_{\rm KL}$              & $D_{KL}^{\hat{p}_1}$     & $D_{KL}^{\hat{p}_2}$    & $D_{KL}^{\hat{p}_3}$    & $D_{KL}^{\hat{p}_4}$ \\
  \hline
  Test A &  $0.22_{-0.04}^{+0.05}$  &  $1.85_{-0.22}^{+0.21}$  &  $0.66_{-0.09}^{+0.11}$  &  $0.11_{-0.03}^{+0.03}$   \\
  Test B &  $0.31_{-0.02}^{+0.03}$  &  $1.79_{-0.16}^{+0.13}$  &  $0.17_{-0.02}^{+0.02}$  &  $0.09_{-0.01}^{+0.02}$   \\
  Test C &  $0.21_{-0.02}^{+0.02}$  &  $1.79_{-0.09}^{+0.09}$  &  $0.55_{-0.05}^{+0.05}$  &  $0.11_{-0.02}^{+0.02}$   \\
  Test D &  $0.18_{-0.04}^{+0.05}$  &  $0.91_{-0.10}^{+0.12}$  &  $1.12_{-0.56}^{+0.22}$  &  $0.42_{-0.07}^{+0.08}$   \\
  Test E &  $0.51_{-0.13}^{+0.14}$  &  $1.35_{-0.19}^{+0.15}$  &  $1.09_{-0.26}^{+0.16}$  &  $0.17_{-0.04}^{+0.04}$   \\
  Test F &  $0.45_{-0.07}^{+0.07}$  &  $2.12_{-0.28}^{+0.28}$  &  $1.73_{-0.35}^{+0.24}$  &  $1.30_{-0.17}^{+0.15}$   \\
  \end{tabular*}
  \label{tb:kldiv}
  {Note. The median and $90\%$ interval are evaluated by repeatedly draw samples from posterior and prior, with each draw gives a KLD value.}
  \end{ruledtabular}
\end{table}

\begin{figure}
  \centering
  \includegraphics[width=0.5\textwidth]{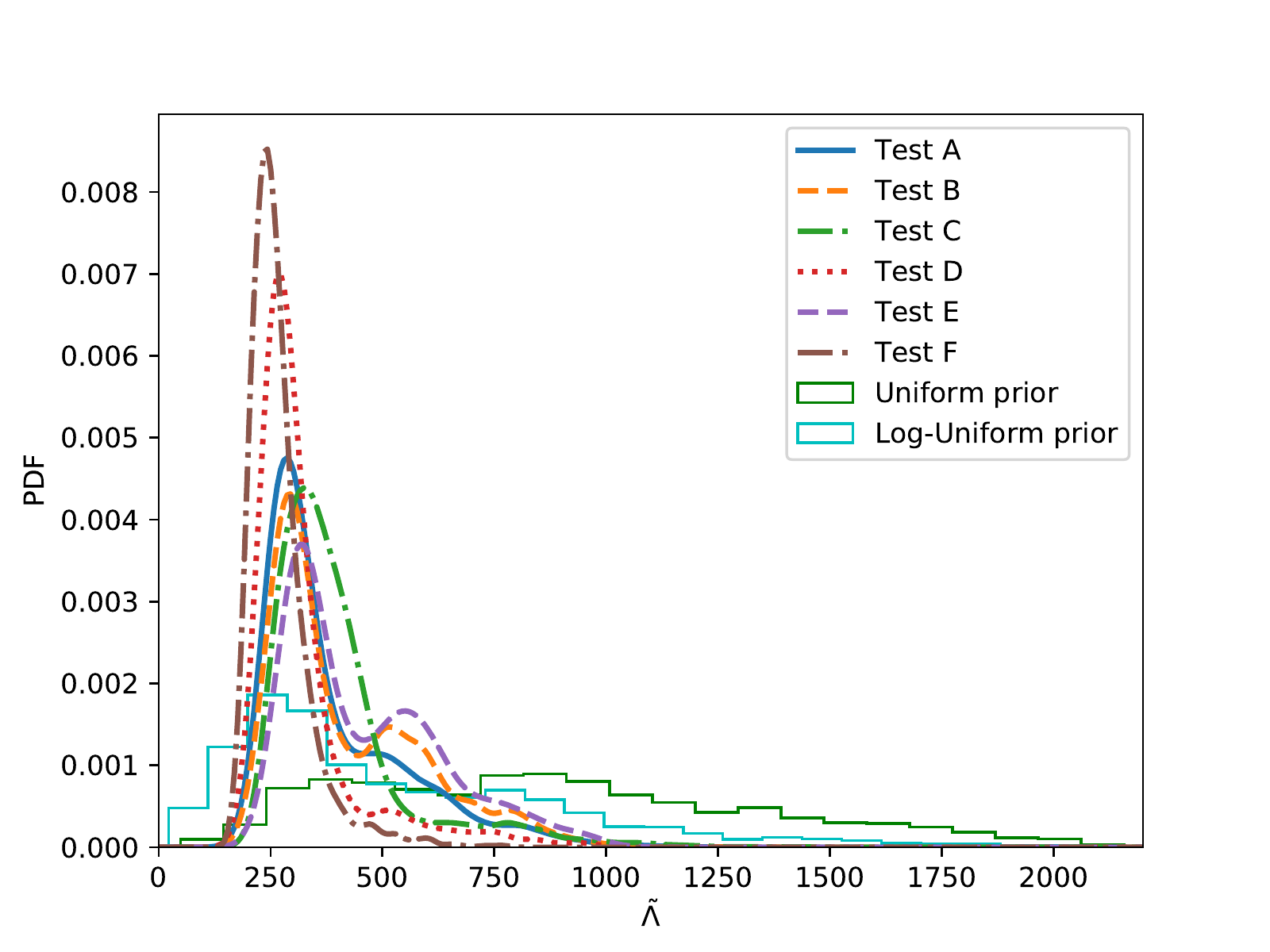}
  \includegraphics[width=0.5\textwidth]{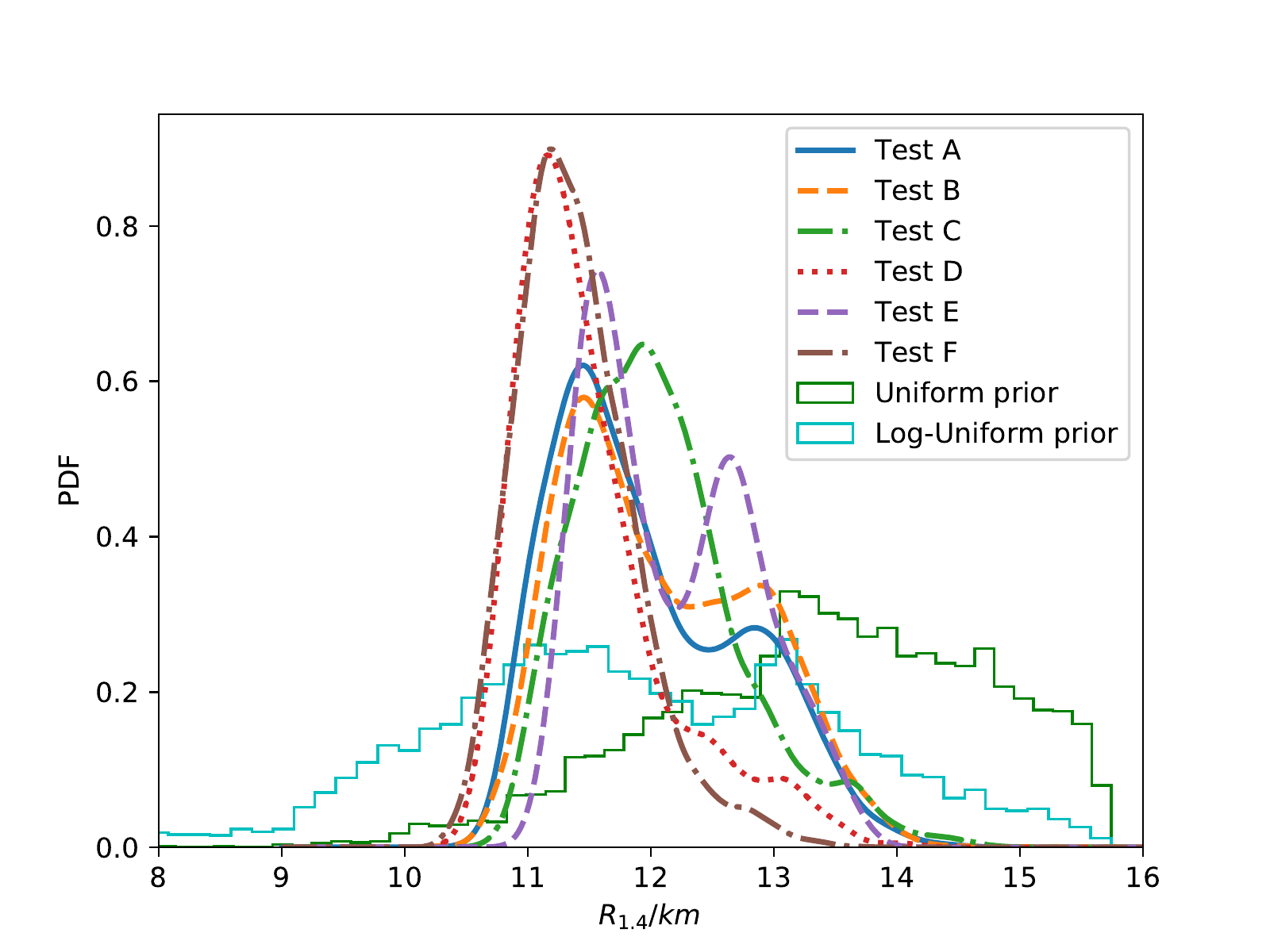}
  \caption{The upper panel shows the dimensionless tidal deformability reconstructed from posterior samples. The lower panel presents the $R_{1.4}$ inferred from each posterior sample.}
  \label{fig:result}
\end{figure}

\begin{figure}
  \centering
  \includegraphics[width=0.5\textwidth]{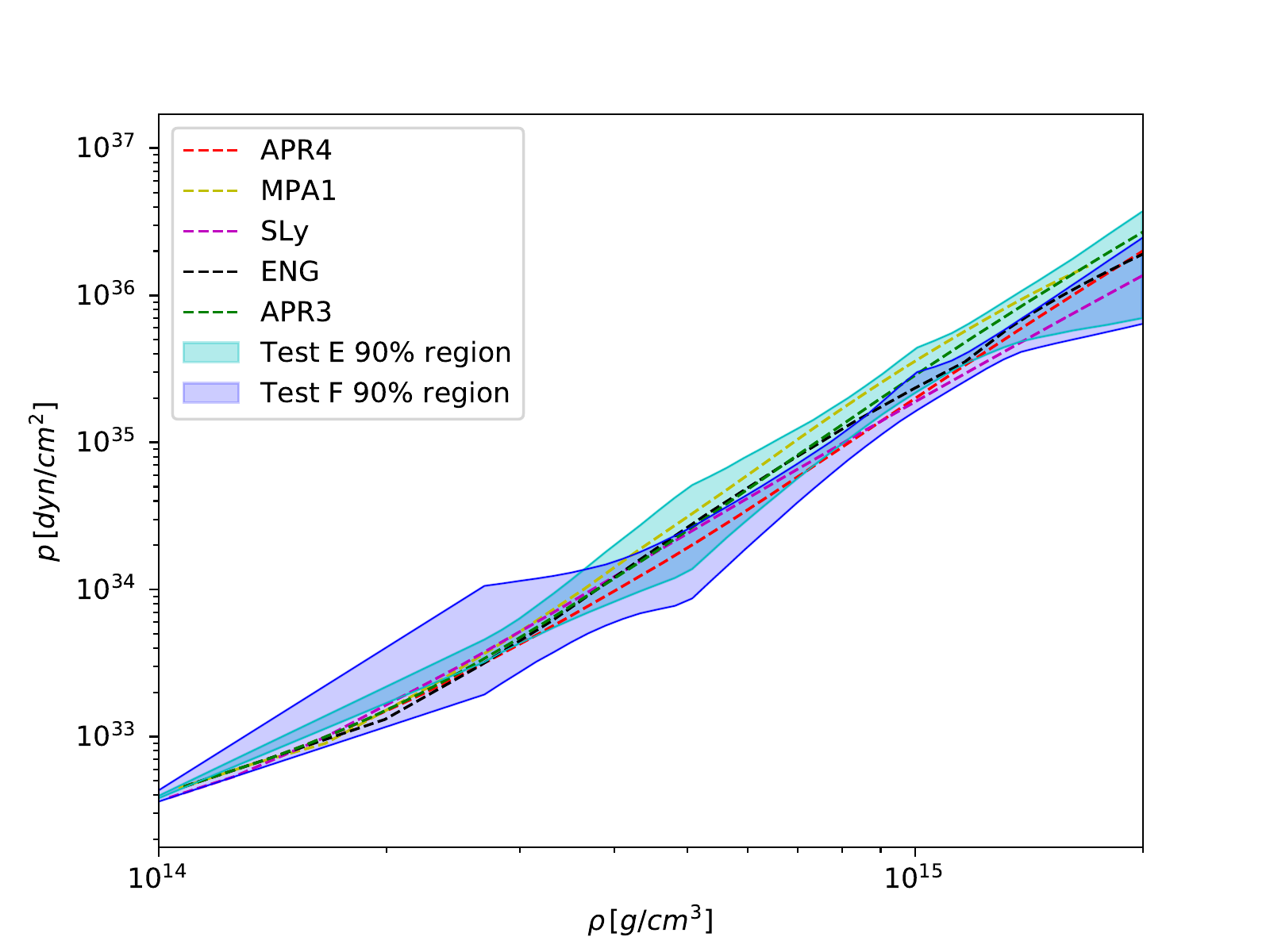}
  \caption{The $90\%$ confidence region of the EoS and some realistic EoSs.}
  \label{fig:conf_region_eos}
\end{figure}

\begin{table*}
  \begin{ruledtabular}
  \caption{$68\%$ and $95\%$  Ranges of Four Pressure Parameters in Different Tests}
  \begin{tabular*}{0.3\textwidth}{lcccccccc}
  Test / Parameter        & $\hat{p}_1$ &        & $\hat{p}_2$  &        & $\hat{p}_3$   &       & $\hat{p}_4$ & \\
                          & $68\%$ & $95\%$      & $68\%$ & $95\%$       & $68\%$ & $95\%$       & $68\%$ & $95\%$      \\
  \hline
  Test A & $5.9_{-3.0}^{+3.4}$ & $5.9_{-4.2}^{+6.5}$ & $1.5_{-0.5}^{+1.4}$ & $1.5_{-0.7}^{+3.3}$ & $3.3_{-0.7}^{+0.8}$ & $3.3_{-1.2}^{+1.4}$ & $2.1_{-0.9}^{+1.2}$ & $2.1_{-1.5}^{+1.8}$ \\
  Test B & $5.5_{-2.6}^{+3.5}$ & $5.5_{-3.8}^{+6.5}$ & $1.7_{-0.7}^{+1.5}$ & $1.7_{-0.9}^{+3.3}$ & $3.8_{-1.3}^{+1.8}$ & $3.8_{-2.0}^{+2.9}$ & $2.2_{-1.0}^{+1.1}$ & $2.2_{-1.6}^{+1.7}$ \\
  Test C & $5.6_{-2.7}^{+3.9}$ & $5.6_{-3.9}^{+6.9}$ & $1.7_{-0.6}^{+0.9}$ & $1.7_{-0.9}^{+3.2}$ & $3.4_{-0.8}^{+0.8}$ & $3.4_{-1.3}^{+1.5}$ & $2.2_{-1.1}^{+1.1}$ & $2.2_{-1.6}^{+1.7}$ \\
  Test D & $3.6_{-1.6}^{+3.3}$ & $3.6_{-2.1}^{+7.4}$ & $1.2_{-0.3}^{+0.8}$ & $1.2_{-0.5}^{+3.1}$ & $3.2_{-0.7}^{+0.9}$ & $3.2_{-1.2}^{+1.7}$ & $1.5_{-0.8}^{+1.2}$ & $1.5_{-1.0}^{+2.2}$ \\
  Test E & $3.9_{-0.5}^{+0.5}$ & $3.9_{-0.7}^{+0.8}$ & $2.2_{-0.8}^{+1.6}$ & $2.2_{-1.0}^{+3.4}$ & $3.2_{-0.7}^{+0.8}$ & $3.2_{-1.1}^{+1.4}$ & $2.0_{-0.9}^{+1.2}$ & $2.0_{-1.4}^{+1.9}$ \\
  Test F & $5.4_{-2.6}^{+3.3}$ & $5.4_{-3.7}^{+6.1}$ & $1.4_{-0.4}^{+0.6}$ & $1.4_{-0.7}^{+1.5}$ & $2.0_{-0.3}^{+0.6}$ & $2.0_{-0.4}^{+1.1}$ & $1.4_{-0.4}^{+0.5}$ & $1.4_{-0.8}^{+1.4}$ \\
  \end{tabular*}
  \label{tb:prop}
  \end{ruledtabular}
\end{table*}


\section{Results}
\label{sec:results}

We calculate Kullback-Leibler divergence (KLD) between prior and posterior following \citet{2019PhRvX...9c1040A} to evaluate how much parameter information is extracted from the data. The KLD between distribution $p$ and $q$ reads
\begin{equation}
    \label{eq:kl_div}
    D_{KL}(p|q) = \int p(x)\log_2(\frac{p(x)}{q(x)})\,dx,
\end{equation}
where $x$ runs over the whole possible range of a parameter. A higher $D_{\rm KL}$ means more parameter information can be extracted from data. In other words, the parameter is well constrained in comparison to the prior.

We can also reconstruct $\tilde{\Lambda}$ using samples of $\{\mathcal{M}_c, q, \hat{p}_1$, $\hat{p}_2, \hat{p}_3, \hat{p}_4\}$ in each Test. This is done by determining $\{M_1, M_2, \Lambda_1, \Lambda_2\}$ from $\{\mathcal{M}_c, q, \hat{p}_1$, $\hat{p}_2, \hat{p}_3, \hat{p}_4\}$ as described in Section \ref{sec:gw}, and then calculate
\[ \tilde{\Lambda} := \frac{16}{13}[\frac{(M_1+12M_2)M_1^4\Lambda_1}{(M_1+M_2)^5}+(1 \leftrightarrow 2)] \].
Similarly, using the posterior samples of EoS parameters $\{ \hat{p}_1$, $\hat{p}_2, \hat{p}_3, \hat{p}_4\}$, we can optimize $h_{\rm c}$ for each single posterior sample to get NS with mass $1.4M_{\odot}$ and then calculate its bulk properties, such as $R_{1.4}$ and $\Lambda_{1.4}$. To avoid the possible bias of bulk properties caused by the prior of the pressure parameters (from the solid green and cyan line in Figure \ref{fig:result}, we can see that this effect is worth noticing), we use the method described in \citet{2019PhRvX...9a1001A} to divide the Kernel Density Estimation (KDE) of the posterior by that of the prior (i.e. to reweight the posterior with prior) and then calculate the $90\%$ highest posterior density (HPD) intervals of bulk properties (see Table \ref{tb:lambdat_r14}).

\subsection{Constraining the EoS}

The gravitational-wave data alone can only constrain $\hat{p}_2$ relatively well (see Test B in Figure \ref{fig:post_pressure} and Table \ref{tb:kldiv}), likely because the information of tidal deformability encoded in the late inspiral state is mostly carried by this parameter \citep{2009PhRvD..79l4033R}.

The relative tight constraints on $M_{\rm TOV}$ help to narrow down the posterior range of $\hat{p}_3$, but have little influence on $\hat{p}_1$, $\hat{p}_2$, and $\hat{p}_4$ (see Figure \ref{fig:post_pressure}). This is also evident in Table \ref{tb:kldiv}. The KLDs of $\hat{p}_1$, $\hat{p}_2$ and $\hat{p}_4$ in Test A and Test B are almost the same, while the KLD of $\hat{p}_3$ changes significantly. This fact indicates that $M_{\rm TOV}$ may be mainly governed by $\hat{p}_3$ rather than other parameters \citep{2009PhRvD..80j3003O, 2009PhRvD..79l4032R}.

We can see from Figure \ref{fig:result} that a tighter mass constraint causes a slight decrease of $\tilde{\Lambda}$ and $R_{1.4}$. Differing from this work, \citet{2018PhRvL.121p1101A} found a strong influence of $M_{\rm TOV}$ on the radius of the NS. Note that \citet{2018PhRvL.121p1101A} adopted a spectral expansion method to parameterize the EoS, while we take a piecewise expansion. The difference in the results may be attributed to the different ways of parameterizing the EoS, as also found in the literature \citep[e.g.][]{2016A&A...591A..25N, 2019arXiv190205078F}. The reason comes partly from the fact that different parameterization methods already give different priors to the global properties of the EoS \citep{2013ApJ...765L...5S}, thus a prior-reweighted posterior of global properties of EoS is needed (see Table \ref{tb:lambdat_r14}).

The results can also be dependent of the waveform model. In comparison to the Test A, the adoption of a different waveform model TaylorF2 (i.e., the Test C) in the analysis leads to a slight decrease of coalescence time (see Figure \ref{fig:gw_pars} in the Appendix \ref{appdx:gw_para}), and a small increase of $\hat{p}_2$ (see Figure \ref{fig:post_pressure}), $\tilde{\Lambda}$, $\Lambda_{1.4}$, and $R_{1.4}$ (see Figure \ref{fig:result} and Table \ref{tb:lambdat_r14}). But no shifts have been observed if the SEOBNRT waveform model is adopted instead.

Changing a flat prior to a log-uniform prior (i.e., Test D) slightly modifies the posterior shapes of $\hat{p}_2$ and $\hat{p}_3$, while the posterior shapes of $\hat{p}_1$ and $\hat{p}_4$ simply follow the prior shapes. This suggests that $\hat{p}_2$ and $\hat{p}_3$ are relatively well constrained by the gravitational-wave data and the maximum mass constraint, but $\hat{p}_1$ and $\hat{p}_4$ are not (see also Table \ref{tb:kldiv}). Also, changing a flat prior to a log-uniform prior leads to a decrease in the posterior of $\tilde{\Lambda}$ and $R_{1.4}$ (see Figure \ref{fig:result}), but we note that after reweighting the posterior with the prior, the HPD range of $\tilde{\Lambda}$ and $R_{1.4}$ in Test A are consistent with those in Test D (see Table \ref{tb:lambdat_r14}). The consistency after re-weighting the posterior with the prior in different analyses that use different priors also provides a proof of the robustness of our results and the importance of reweighting a posterior with a prior.

The addition nuclear constraints (i.e., Test E) sharply narrows down the prior range of $\hat{p}_1$, and boosts $\hat{p}_2$ to a slightly higher value, but does not help constraining $\hat{p}_3$ and $\hat{p}_4$ (see Figure \ref{fig:post_pressure} and Table \ref{tb:prop}). Additionally, the inclusion of the nuclear constraints favors larger $\tilde{\Lambda}$, $R_{1.4}$, and $\Lambda_{1.4}$ than Test A (see Figure \ref{fig:result} and Table \ref{tb:lambdat_r14}).

The joint constraints on the EoS by the LMXB sources with thermonuclear bursts and gravitational-wave data (i.e., Test F) can well constrain $\hat{p}_2$, $\hat{p}_3$, and $\hat{p}_4$ (see Figure \ref{fig:post_pressure} and Table \ref{tb:kldiv}). The constraint on $\hat{p}_2$ mainly comes from radius measurement of these sources, because the radius of an NS is mainly determined by $\hat{p}_2$ \citep{2001ApJ...550..426L}. Our $R_{1.4}$ (see Table \ref{tb:lambdat_r14}) is larger than that of \citet{2013ApJ...772....7G}; the difference may come from the fact that they use the quiescent LMXB data, but we use the data of LMXB sources with thermonuclear bursts.

In particular, small $R_{1.4}$ and $\Lambda_{1.4}$ are produced \citep[see also][]{2019arXiv190205078F} in Test F, differing from what we have found in Test E, and indicating that the nuclear data and the $M/R$ measurement of LMXB sources that have thermonuclear bursts may not be fully consistent with each other. Such a tension may be resolved in the future as long as the nuclear data can be better measured/understood and the measurements of the NSs in the LMXB sources with thermonuclear bursts have been significantly improved (so far, the measured radii can still suffer from serious systematic uncertainties).

Our gravitational-wave parameters are nicely in agreement with \citet{2019PhRvX...9a1001A} in all six tests (see the Appendix \ref{appdx:gw_para}). Additionally, as shown in Figure \ref{fig:result} and Table \ref{tb:lambdat_r14}, for both Test A and Test C, the resulting $\tilde{\Lambda}$ are also consistent with \citet{2019PhRvX...9a1001A}.

With the posterior of pressure parameters, it is straightforward to calculate the allowed region of the EoS. As shown in Figure \ref{fig:conf_region_eos}, in comparison to the default scenario (Test A), Test F can improve the constraints in the high-density region while Test E can better constrain the low-density region. We also compare these results with some realistic EoSs (see the web reference in footnote 2) and find reasonable agreement. Since the start of the O3 run of advanced LIGO/Virgo in 2019 April, a few NS merger gravitational-wave events have been reported \footnote{https://gracedb.ligo.org/latest/}. The release of these new data is expected to significantly improve the constraints on the EoS of NSs.

\begin{table}
  \begin{ruledtabular}
  \caption{$90\%$ HPD Range of a Prior-reweighted Posterior of $R_{1.4}$, $\Lambda_{1.4}$ and $\tilde{\Lambda}$ Directly Inferred from Posterior Samples of $\{\hat{p_1}, \hat{p_2}, \hat{p_3}, \hat{p_4}\}$}
  \begin{tabular*}{0.3\textwidth}{lccc}
  Test/Property             &   $R_{1.4}/km$   & $\Lambda_{1.4}$  & $\tilde{\Lambda}$ \\
  \hline
  Test A  &  $11.5_{-0.8}^{+1.5}$  &  $300_{-130}^{+300}$  &  $310_{-160}^{+320}$  \\
  Test B  &  $11.5_{-0.9}^{+1.5}$  &  $330_{-170}^{+310}$  &  $340_{-190}^{+350}$  \\
  Test C  &  $11.7_{-0.9}^{+1.0}$  &  $320_{-160}^{+280}$  &  $350_{-200}^{+270}$  \\
  Test D  &  $11.4_{-0.8}^{+1.3}$  &  $290_{-140}^{+310}$  &  $310_{-180}^{+340}$  \\
  Test E  &  $11.8_{-0.7}^{+1.2}$  &  $390_{-210}^{+280}$  &  $400_{-230}^{+310}$  \\
  Test F  &  $11.1_{-0.6}^{+0.7}$  &  $220_{-90}^{+90}$    &  $260_{-190}^{+90}$  \\
  \end{tabular*}
  \label{tb:lambdat_r14}
  \end{ruledtabular}
\end{table}

\subsection{Constraining the properties of galactic double NS systems}

With the posterior samples of $\{M_1,M_2,\Lambda_1,\Lambda_2\}$ of GW170817 and some ``universal" relations, it is possible to ``extrapolate" these properties to similar masses and thus get the constraints of some global properties at these given masses, in particular those accurately measured for the galactic double NS systems \citep{2018ApJ...868L..22L,2019PhRvD..99l3026K}.

\begin{figure}
  \centering
  \includegraphics[width=0.46\textwidth]{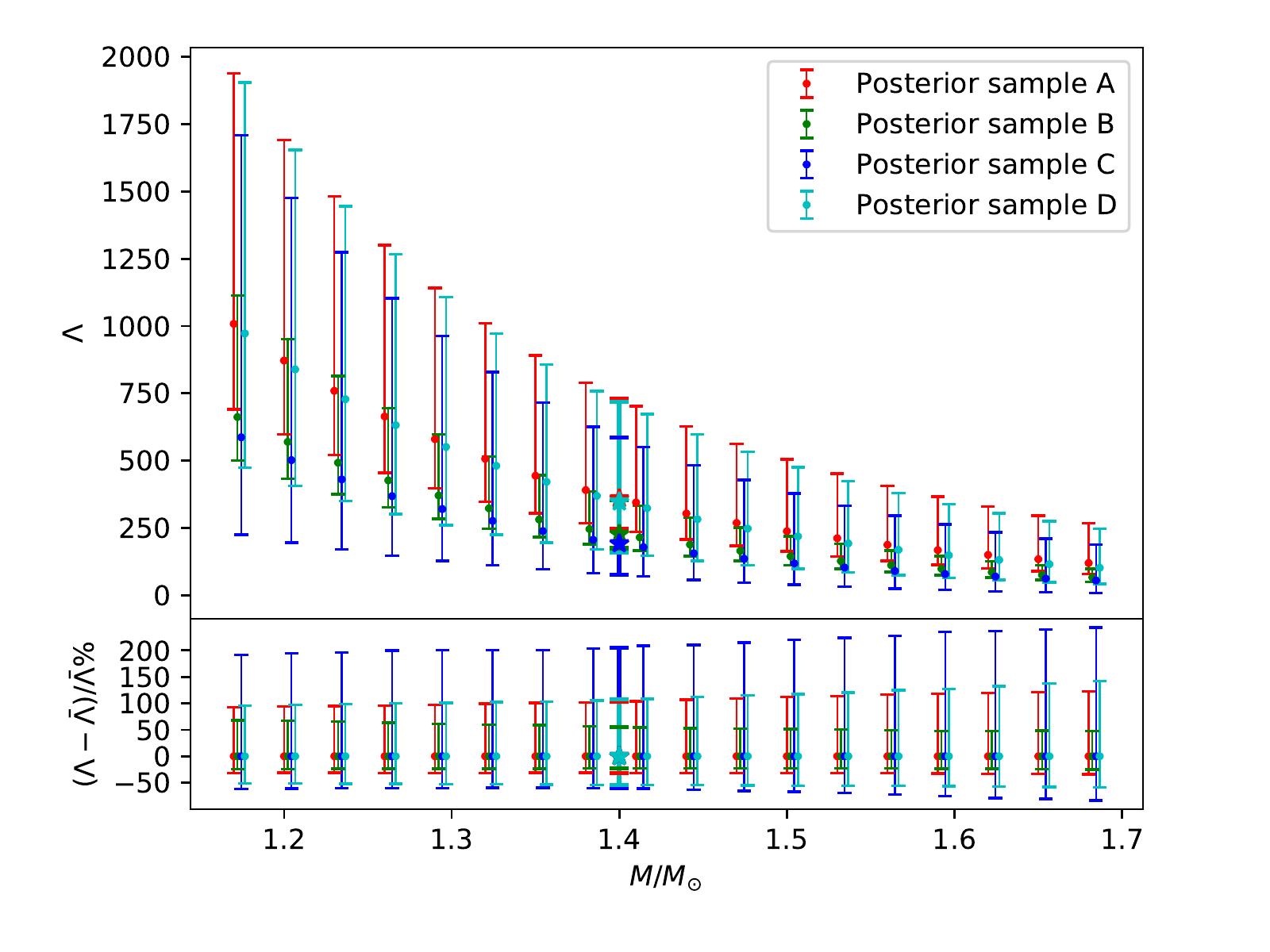}
  \includegraphics[width=0.46\textwidth]{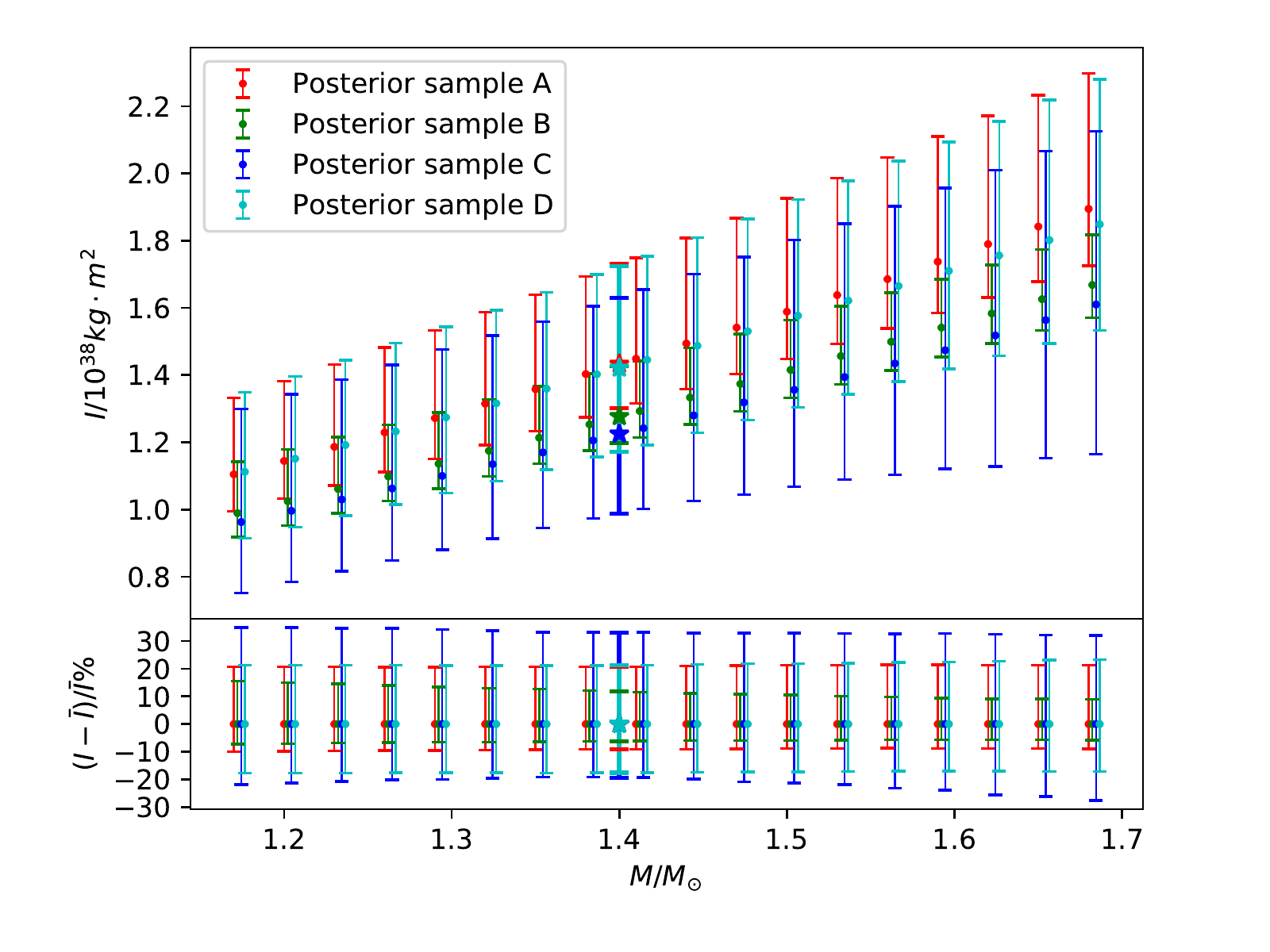}
  \includegraphics[width=0.46\textwidth]{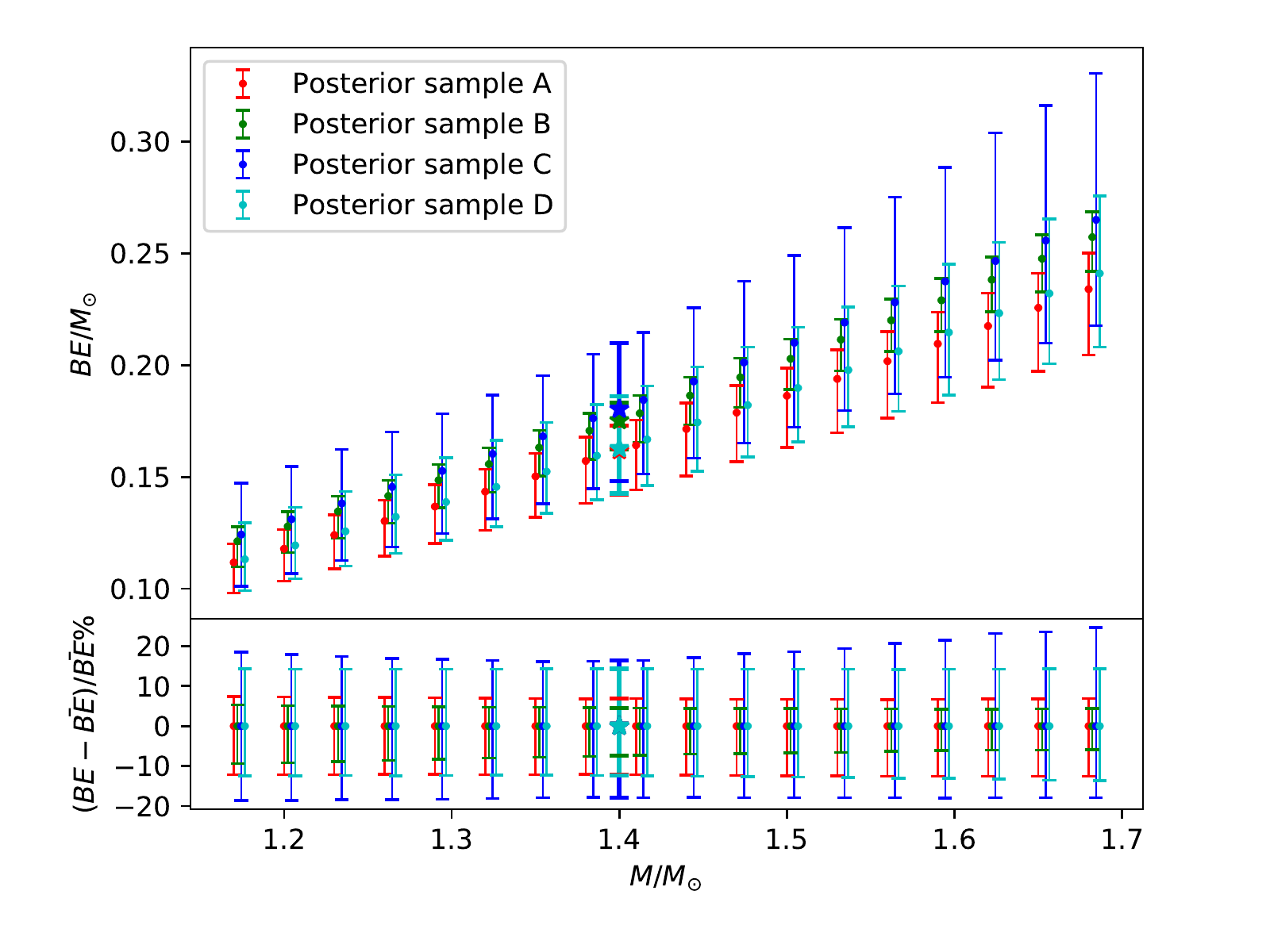}
  \caption{Upper panel: $90\%$ interval of dimensionless tidal deformability and its relative error. Middle panel: $90\%$ interval of moment of inertia and its relative error. Lower panel: $90\%$ interval of binding energy and its relative error. These intervals are all calculated from different groups of posterior samples and universal relations at each mass. Here the red, green, blue, and cyan error bars respectively show the $90\%$ intervals for Posterior Samples A, B, C and D. The stars show the constraint at the canonical mass $M=1.4M_{\odot}$.}
  \label{fig:I-L-BE}
\end{figure}

\begin{table}
  \begin{ruledtabular}
  \caption{$90\%$ Interval of Dimensionless Tidal Deformability, Moment of Inertia and Binding Energy of an NS with Gravitational Mass $1.4M_{\odot}$}
  \begin{tabular*}{0.5\textwidth}{lccc}
  Samples   & $\Lambda_{1.4}$     & $I_{1.4}/10^{38} \rm kg \cdot m^2$    & BE$_{1.4}/M_{\odot}$  \\
  \hline
  Sample A &  ${360}^{+370}_{-110}$  &  ${1.43}^{+0.30}_{-0.13}$  &  ${0.16}^{+0.01}_{-0.02}$  \\
  Sample B &  ${230}^{+130}_{-50}$   &  ${1.28}^{+0.15}_{-0.08}$  &  ${0.18}^{+0.01}_{-0.01}$  \\
  Sample C &  ${190}^{+390}_{-120}$  &  ${1.23}^{+0.41}_{-0.24}$  &  ${0.18}^{+0.03}_{-0.03}$  \\
  Sample D &  ${350}^{+370}_{-190}$  &  ${1.42}^{+0.30}_{-0.25}$  &  ${0.16}^{+0.02}_{-0.02}$  \\
  \end{tabular*}
  \label{tb:ILBE14}
  {Note. These properties are inferred from posterior samples of $\{M_1, M_2, \Lambda_1, \Lambda_2 \}$ and universal relations described in the main text.}
  \end{ruledtabular}
\end{table}

\subsubsection{Constraint method}
\label{sec:cons-method}

Tidal deformability can be expanded into a Taylor series around a ``canonical" reference mass $M_{\rm ref}$ \citep{2013PhRvL.111g1101D}. Below we adopt a linear expansion following \citet{2013PhRvL.111g1101D} and \citet{2018PhRvL.121p1101A}
\begin{equation}
    \label{eq:l-expand}
    \lambda{(M)} \simeq \lambda_{\rm ref} + \lambda^{1}(M-M_{\rm ref})/M_{\odot},
\end{equation}
where $\lambda(M) \equiv \Lambda(M) (GM/c^2)^5$ is the tidal deformability of the NS with a gravitational mass $M$, $\Lambda$ is its dimensionless form, and $G$ is Newton's gravitational constant. For a given reference mass and a single posterior sample of ($M_1, M_2, \Lambda_1, \Lambda_2$), we can solve equation (\ref{eq:l-expand}) to get a unique $\lambda_{\rm ref}$, $\lambda^{1}$, and then $\Lambda_{\rm ref}$. With a group of posterior samples of ($M_1, M_2, \Lambda_1, \Lambda_2$), the distribution of $\Lambda_{\rm ref}$ can be inferred at this reference mass. After varying the reference mass in a given range and following the same procedure outlined above, we can get the corresponding constraints on dimensionless tidal deformability of NSs in this mass range (see the upper panel of Figure \ref{fig:I-L-BE}).

The NS's tidal deformability $\Lambda$ and dimensionless moment of inertia $\bar{I} \equiv c^4I/G^2M^3$ are found to have an EoS-insensitive relation, which is the so-called I-Love relation, where $I$ is the moment of inertia. Here we take the function form from \citet{2013Sci...341..365Y}, which reads
\begin{equation}
    \label{eq:IL}
    \log_{10}{\bar{I}} = \sum_{n=0}^{4}a_{n}(\log_{10}{\Lambda})^n,
\end{equation}
where $a_{n}$ are the fit coefficients, which are adopted from \citet{2018ApJ...868L..22L}. For a reference mass $M_{\rm ref}$, a group of possible $\Lambda_{\rm ref}$ are calculated from equation (\ref{eq:l-expand}) for each group of a posterior sample, then we can calculate a group of possible moment of inertia $I$ from equation (\ref{eq:IL}). After varying the reference mass, we get constraints on the moments of inertia (see middle panel of Figure \ref{fig:I-L-BE}).

The BE also have an EoS-insensitive relation with dimensionless tidal deformability $\bar{I}$, i.e.,
\begin{equation}
    \label{eq:BE}
    BE/M = \sum_{n=0}^{4} b_{n}\bar{I}^{-n}.
\end{equation}
Here we take the fit coefficients $b_{n}$ from \citet{2016EPJA...52...18S} and use equations (\ref{eq:l-expand}$-$\ref{eq:BE}) to calculate a sample of BE and its $90\%$ range--for a given reference mass and a given posterior sample, then change reference mass and repeat the same procedure to set constraints on the whole mass range considered (see the lower panel of Figure \ref{fig:I-L-BE}).

\subsubsection{Posterior choices and mass range}

For our current purposes we adopt four posterior samples, including two obtained in this work (i.e., the Posterior Sample A corresponding to that of Test E and the Posterior Sample B for Test F) and the other two adopted from \citet{2018PhRvL.121p1101A}.  The Posterior Sample C is available on the web\footnote{https://dcc.ligo.org/LIGO-P1800115/public}, which is the result of a universal relation-based analysis. The Posterior Sample D is taken from the same literature but it is the result of a spectral EOS parameterization analysis that imposes a maximum gravitational mass of at least $1.97M_\odot$ \citep{2013Sci...340..448A}.

Here we focus on the NS masses between $1.17~M_\odot$ and $1.68~M_{\odot}$, which cover the most probable mass range of galactic double NS systems. Please note that here we take the $68\%$ lower (upper) limit of the lowest (highest) mass in 12 galactic double NS systems whose individual masses were accurately measured \citep[see][and the references therein]{2019ApJ...876...18F}.

\subsubsection{Constraint results}

The resulting $\Lambda$, $I$, and BE in the mass range of galactic double neutron stars are summarized in Figure \ref{fig:I-L-BE}. We can see that the constraints in all the four scenarios are consistent with each other (see Table \ref{tb:ILBE14}), giving the rather large uncertainties. However, there are some interesting general tendencies. Posterior Sample A and Posterior Sample D tend to favor higher $\Lambda$ and $I$, but have a lower BE than the cases of Posterior Sample B and Posterior Sample C. Interestingly, similar conclusions about the difference of $\Lambda_{1.4}$ in Posterior Sample C and Posterior Sample D were drawn in \citet{2018PhRvL.121p1101A}, who attributed the difference to the additional $M_{\rm TOV}$ constraint. Besides, although dimensionless tidal deformability decreases very quickly with the increasing reference mass (see Figure \ref{fig:I-L-BE}), the moment of inertia and the binding energy increase almost linearly with $M_{\rm ref}$.  In the meantime, the lower error is smaller than the upper error in all these cases. Additionally, because of the use of universal relations, the relative error of $I$ is significantly smaller than that of $\Lambda$, and the relative error of BE is systematically smaller than that of $I$.

It is also evident from Figure \ref{fig:I-L-BE} that the more constraints/data we add, the smaller relative error of the inferred global properties of galactic NSs we get. Posterior Sample D infers a smaller relative error of global properties than that of Posterior Sample C, because the former adopts an additional mass constraint $M_{\rm TOV}>1.97M_{\odot}$. Posterior Sample A adopts a tighter mass constraint $2.5M_{\odot}>M_{\rm TOV}>2.06M_{\odot}$ and additional nuclear constraint than Posterior Sample D, so it gets smaller relative errors of $\Lambda$, $I$, and BE than the latter. Interestingly, Posterior Sample B gets the smallest relative error because of the additional tighter mass constraint, and because the data of LMXB sources with burst are considered.

The canonical global properties of Posterior Samples A and B shown in table \ref{tb:ILBE14} are consistent with those of Test E and F shown in Table \ref{tb:lambdat_r14}, because they adopt the same posterior sample, with the latter been directly reconstructed from posterior sample of EoS parameters $\{\hat{p_1}, \hat{p_2}, \hat{p_3}, \hat{p_4}\}$ and the former being inferred from NS properties $\{M_1, M_2, \Lambda_1, \Lambda_2\}$ of sources of GW170817 and universal relations.

\section{Discussion and Summary}
\label{sec:sum_cons}

We combine the gravitational-wave data, LMXB sources with thermonuclear bursts or nuclear constraints on the symmetry energy together to do the joint analysis, finding that different data sets can constrain different pressure parameters, i.e., the constraint on $\hat{p}_1$ mainly comes from nuclear constraints, the constraint on $\hat{p}_2$ is mainly contributed by the gravitational-wave data and the LMXB sources with thermonuclear bursts, the constraint on $\hat{p}_3$ heavily relies on the LMXB source data and the current bounds of $M_{\rm TOV}$, the range of $\hat{p}_4$ can be slightly narrowed down by LMXB sources with thermonuclear bursts. We also find that nuclear constraints tend to give larger $R_{1.4}$ and LMXB sources with thermonuclear bursts tend to indicate smaller ones. Our $\Lambda_{1.4}$ bounds found in Test F are consistent with those of \citet{2019PhRvD..99l3026K}. However, our median value is a bit higher than that found in \citet{2019PhRvD..99l3026K}. Such a difference is likely caused by the very different analysis methods and by our additional bounds on $M_{\rm TOV}$.

With some EoS-insensitive relations and our posterior samples, we have evaluated the possible ranges of tidal deformability, moment of inertia, and BE of NSs in the mass range of galactic double NS systems. The constraints in all the four scenarios are consistently (see Table \ref{tb:ILBE14}) produce rather high uncertainties. Particularly, for the NS with a canonical mass of $1.4M_\odot$, we have $I_{1.4} = {1.43}^{+0.30}_{-0.13} \times 10^{38}~{\rm kg \cdot m^2}$, $\Lambda_{1.4} = 390_{-210}^{+280}$, $R_{1.4} = 11.8_{-0.7}^{+1.2}~{\rm km}$, and $BE_{1.4} = {0.16}^{+0.01}_{-0.02} M_{\odot}$  if the constraints from the nuclear data and the gravitational-wave data have been considered together. For the joint analysis of gravitational-wave data and LMXB sources with thermonuclear bursts, we have $I_{1.4} = {1.28}^{+0.15}_{-0.08} \times 10^{38}~{\rm kg \cdot m^2}$, $\Lambda_{1.4} = 220_{-90}^{+90}$, $R_{1.4} = 11.1_{-0.6}^{+0.7}~{\rm km}$ and $BE_{1.4} = {0.18}^{+0.01}_{-0.01} M_{\odot}$. These results suggest that the current constraints on $\Lambda$ still suffer from significant systematic uncertainties \citep[see also, e.g.][]{2013ApJ...771...51L,2018PhRvL.121p1101A}, while  $I_{1.4}$ and $BE_{1.4}$ are relatively better bounded.

\acknowledgments
We thank the anonymous referee for the helpful suggestions. This work was supported in part by NSFC under grants of No. 11525313 (i.e., Funds for Distinguished Young Scholars), No. 11433009 and No. 11773078, the Funds for Distinguished Young Scholars of Jiangsu Province (No. BK20180050), the Chinese Academy of Sciences via the Strategic Priority Research Program (grant No. XDB23040000), and the Key Research Program of Frontier Sciences (No. QYZDJ-SSW-SYS024).

\software{Bilby \citep[version 0.5.5, ascl:1901.011, \url{https://git.ligo.org/lscsoft/bilby/}]{2019ascl.soft01011A}, PyCBC \citep[version 1.13.6, ascl:1805.030, \url{http://doi.org/10.5281/zenodo.3265452}]{2018ascl.soft05030T}, PyMultiNest \citep[version 2.6, ascl:1606.005, \url{https://github.com/JohannesBuchner/PyMultiNest}]{2016ascl.soft06005B}.}

\appendix

\begin{figure*}
  \centering
  \includegraphics[width=1\textwidth]{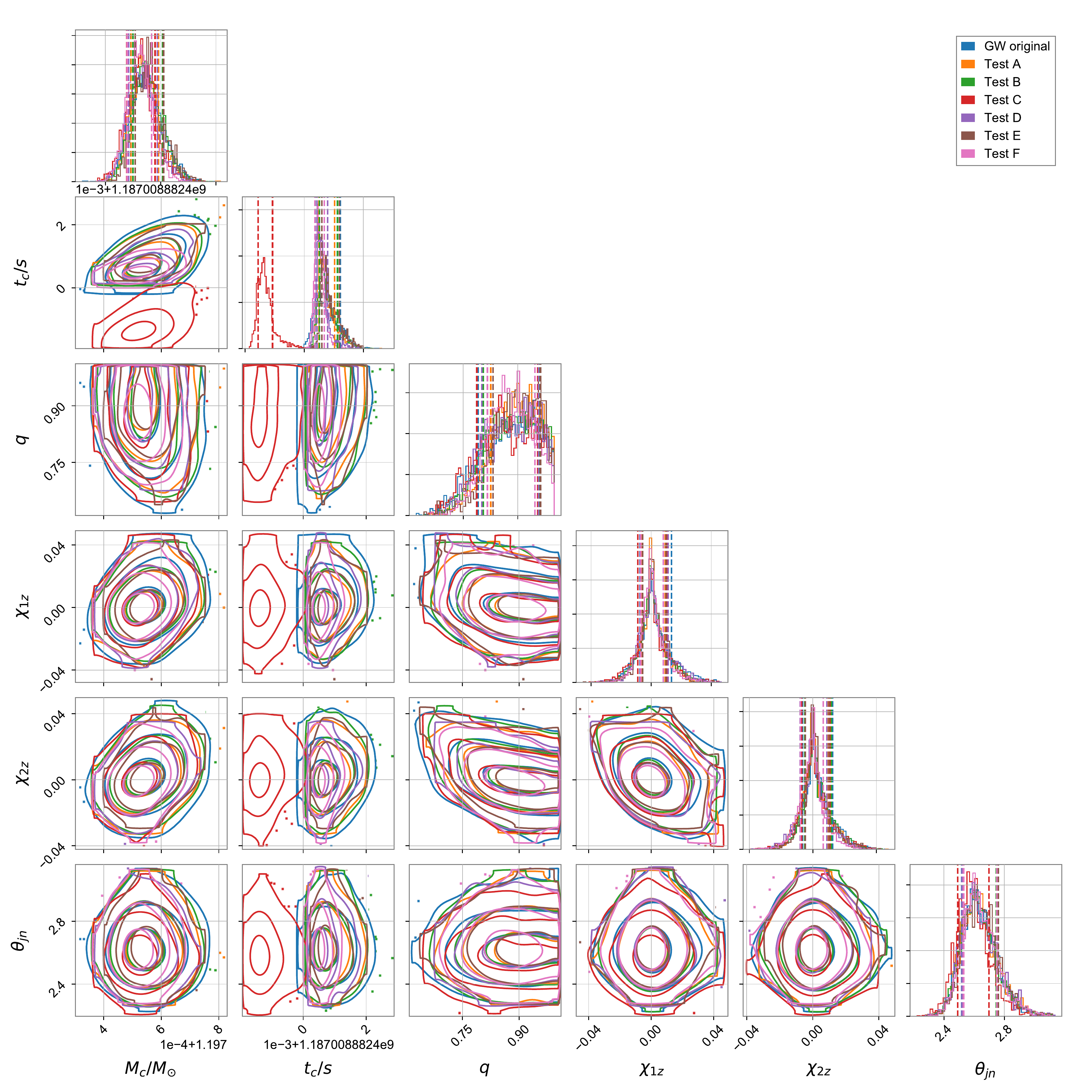}
  \caption{Posterior distribution of gravitational-wave parameters in each test.}
  \label{fig:gw_pars}
\end{figure*}

\begin{table*}
  \begin{ruledtabular}
  \caption{$90\%$ Interval of Gravitational-wave Detection Frame Parameters in Different Tests}
  \begin{tabular*}{0.3\textwidth}{lcccccc}
  Test / Parameter        & $\mathcal{M}_c/M_{\odot}$ & $t_c/s$    & $q$       & $\chi_{1z}$      & $\chi_{2z}$   & $\theta_{jn}/deg$  \\
  \hline
  GW original & $1.1975_{-0.0001}^{+0.0001}$ & $1187008882.4307_{-0.0005}^{+0.0008}$ & $0.88_{-0.14}^{+0.10}$ & $0.00_{-0.01}^{+0.02}$ & $0.00_{-0.02}^{+0.02}$ & $150_{-9}^{+12}$ \\
  Test A & $1.1975_{-0.0001}^{+0.0001}$ & $1187008882.4307_{-0.0003}^{+0.0007}$ & $0.90_{-0.12}^{+0.09}$ & $0.00_{-0.01}^{+0.02}$ & $0.00_{-0.01}^{+0.02}$ & $151_{-9}^{+13}$ \\
  Test B & $1.1975_{-0.0001}^{+0.0001}$ & $1187008882.4307_{-0.0003}^{+0.0007}$ & $0.89_{-0.13}^{+0.10}$ & $0.00_{-0.01}^{+0.02}$ & $0.00_{-0.01}^{+0.02}$ & $150_{-9}^{+14}$ \\
  Test C & $1.1975_{-0.0001}^{+0.0001}$ & $1187008882.4286_{-0.0003}^{+0.0007}$ & $0.87_{-0.13}^{+0.11}$ & $0.00_{-0.01}^{+0.02}$ & $0.00_{-0.02}^{+0.02}$ & $148_{-9}^{+12}$ \\
  Test D & $1.1975_{-0.0001}^{+0.0001}$ & $1187008882.4306_{-0.0002}^{+0.0006}$ & $0.88_{-0.12}^{+0.10}$ & $0.00_{-0.01}^{+0.02}$ & $0.00_{-0.01}^{+0.02}$ & $150_{-9}^{+13}$ \\
  Test E & $1.1976_{-0.0001}^{+0.0001}$ & $1187008882.4308_{-0.0003}^{+0.0007}$ & $0.90_{-0.12}^{+0.08}$ & $0.00_{-0.01}^{+0.02}$ & $0.00_{-0.01}^{+0.02}$ & $151_{-9}^{+12}$ \\
  Test F & $1.1975_{-0.0001}^{+0.0001}$ & $1187008882.4305_{-0.0002}^{+0.0003}$ & $0.88_{-0.10}^{+0.10}$ & $-0.00_{-0.01}^{+0.01}$ & $-0.00_{-0.01}^{+0.02}$ & $151_{-9}^{+13}$ \\
  \end{tabular*}
  \label{tb:gw_prop}
  \end{ruledtabular}
  {Note. The unit of $\theta_{jn}$ is transformed from rad to degree to compare with other analyses.}

\end{table*}


\section{Gravitational-wave Parameters}
\label{appdx:gw_para}

To further check our results, we carry out another test labeled as ``GW", which samples $\{\mathcal{M}_c, q, \Lambda_1, \Lambda_2, \chi_1, \chi_2,$ $ \theta_{jn}, t_c, \Psi\}$. We have also calculated the properties of gravitational-wave parameters (see Table \ref{tb:gw_prop} and Figure \ref{fig:gw_pars}). Our results are self-consistent among all test scenarios and are in agreement with \citet{2019PhRvX...9a1001A}. We do not provide the property of polarization $\Psi$, because it is poorly constrained and carries little astrophysical information. The error of $\theta_{jn}$ is reduced compared with that of \citet{2017PhRvL.119p1101A} but consistent with \citet{2019PhRvX...9a1001A}, because the sky location of GW170817 is fixed to its optical counterpart. The coalescence time $t_c$ of Test C is slightly different from other tests because of the adoption of the TaylorF2 waveform model. In our analysis the ``aligned spin prior" is assumed, which implies that the information in the direction that is perpendicular to the orbital angular momentum has been lost. In other words, we are simply constraining the spin in the $z$ direction, for which a zero median value is expected \citep[see also][]{2019PhRvX...9a1001A}.


\begin{thebibliography}

\bibitem[Abbott et al.(2017a)]{2017PhRvL.119p1101A} Abbott, B.~P., Abbott, R., Abbott, T.~D., et al.\ 2017, \prl, 119, 161101

\bibitem[Abbott et al.(2017b)]{2017ApJ...848L..12A} Abbott, B.~P., Abbott, R., Abbott, T.~D., et al.\ 2017, \apjl, 848, L12

\bibitem[Abbott et al.(2018)]{2018PhRvL.121p1101A} Abbott, B.~P., Abbott, R., Abbott, T.~D., et al.\ 2018, \prl, 121, 161101

\bibitem[Abbott et al.(2019)]{2019PhRvX...9a1001A} Abbott, B.~P., Abbott, R., Abbott, T.~D., et al.\ 2019, Physical Review X, 9, 011001

\bibitem[Abbott et al.(2019)]{2019PhRvX...9c1040A} Abbott, B.~P., Abbott, R., Abbott, T.~D., et al.\ 2019, Physical Review X, 9, 031040

\bibitem[Akmal et al.(1998)]{1998PhRvC..58.1804A} Akmal, A., Pandharipande, V.~R., \& Ravenhall, D.~G.\ 1998, \prc, 58, 1804

\bibitem[Allen et al.(2012)]{2012PhRvD..85l2006A} Allen, B., Anderson, W.~G., Brady, P.~R., et al.\ 2012, \prd, 85, 122006

\bibitem[Annala et al.(2018)]{2018PhRvL.120q2703A} Annala, E., Gorda, T., Kurkela, A., \& Vuorinen, A.\ 2018, \prl, 120, 172703

\bibitem[Antoniadis et al.(2013)]{2013Sci...340..448A} Antoniadis, J., Freire, P.~C.~C., Wex, N., et al.\ 2013, Science, 340, 448

\bibitem[Ashton et al.(2019)]{2019ascl.soft01011A} Ashton, G., H{\"u}bner, M., Lasky, P.~D., et al.\ 2019, Bilby: Bayesian inference library, ascl:1901.011

\bibitem[Baillot d'Etivaux et al.(2019)]{2019arXiv190501081B} Baillot d'Etivaux, N., Guillot, S., Margueron, J., et al.\ 2019, arXiv:1905.01081

\bibitem[Biwer et al.(2019)]{2019PASP..131b4503B} Biwer, C.~M., Capano, C.~D., De, S., et al.\ 2019, \pasp, 131, 024503

\bibitem[Buchner(2016)]{2016ascl.soft06005B} Buchner, J.\ 2016, PyMultiNest: Python interface for MultiNest, ascl:1606.005

\bibitem[Breu \& Rezzolla (2016)]{2016MNRAS.459..646B} Breu, C., \& Rezzolla, L., 2016, MNRAS, 459, 646

\bibitem[Cromartie et al.(2019)]{2019NatAs.tmp..439C} Cromartie, H.~T., Fonseca, E., Ransom, S.~M., et al.\ 2019, Nature Astronomy

\bibitem[De et al.(2018)]{2018PhRvL.121i1102D} De, S., Finstad, D., Lattimer, J.~M., et al.\ 2018, \prl, 121, 091102

\bibitem[Del Pozzo et al.(2013)]{2013PhRvL.111g1101D} Del Pozzo, W., Li, T.~G.~F., Agathos, M., Van Den Broeck, C., \& Vitale, S.\ 2013, \prl, 111, 071101

\bibitem[Demorest et al.(2010)]{2010Natur.467.1081D} Demorest, P.~B., Pennucci, T., Ransom, S.~M., Roberts, M.~S.~E., \& Hessels, J.~W.~T.\ 2010, \nat, 467, 1081

\bibitem[Fan et al.(2013)]{Fan2013} Fan, Y. Z., Wu, X. F., \& Wei, D. M. 2013, PhRvD, 88, 067304

\bibitem[Farrow et al.(2019)]{2019ApJ...876...18F} Farrow, N., Zhu, X.-J., \& Thrane, E.\ 2019, \apj, 876, 18

\bibitem[Fasano et al.(2019)]{2019arXiv190205078F} Fasano, M., Abdelsalhin, T., Maselli, A., \& Ferrari, V.\ 2019, \prl, 123, 141101

\bibitem[Fattoyev et al.(2018)]{2018PhRvL.120q2702F} Fattoyev, F.~J., Piekarewicz, J., \& Horowitz, C.~J.\ 2018, \prl, 120, 172702

\bibitem[Guillot et al.(2013)]{2013ApJ...772....7G} Guillot, S., Servillat, M., Webb, N.~A., \& Rutledge, R.~E.\ 2013, \apj, 772, 7

\bibitem[Kumar, \& Landry(2019)]{2019PhRvD..99l3026K} Kumar, B., \& Landry, P.\ 2019, \prd, 99, 123026

\bibitem[Kurkela et al.(2014)]{2014ApJ...789..127K} Kurkela, A., Fraga, E.~S., Schaffner-Bielich, J., \& Vuorinen, A.\ 2014, \apj, 789, 127

\bibitem[Krastev, \& Li(2019)]{2019JPhG...46g4001K} Krastev, P.~G., \& Li, B.-A.\ 2019, Journal of Physics G Nuclear Physics, 46, 074001

\bibitem[Landry \& Kumar(2018)]{2018ApJ...868L..22L} Landry, P., \& Kumar, B.\ 2018, \apjl, 868, L22

\bibitem[Lattimer \& Prakash(2001)]{2001ApJ...550..426L} Lattimer, J.~M., \& Prakash, M.\ 2001, \apj, 550, 426

\bibitem[Lattimer(2012)]{2012ARNPS..62..485L} Lattimer, J.~M.\ 2012, ARNPS, 62, 485

\bibitem[Lattimer \& Lim(2013)]{2013ApJ...771...51L} Lattimer, J.~M., \& Lim, Y.\ 2013, \apj, 771, 51

\bibitem[Lattimer \& Steiner(2014a)]{2014EPJA...50...40L} Lattimer, J.~M., \& Steiner, A.~W.\ 2014, EPJA, 50, 40

\bibitem[Lattimer \& Steiner(2014b)]{2014ApJ...784..123L} Lattimer, J.~M., \& Steiner, A.~W.\ 2014, \apj, 784, 123

\bibitem[Lattimer \& Prakash(2016)]{2016PhR...621..127L} Lattimer, J.~M., \& Prakash, M.\ 2016, \physrep, 621, 127

\bibitem[Levan et al.(2017)]{2017ApJ...848L..28L} Levan, A.~J., Lyman, J.~D., Tanvir, N.~R., et al.\ 2017, \apjl, 848, L28

\bibitem[Lim \& Holt(2018)]{2018PhRvL.121f2701L} Lim, Y., \& Holt, J.~W.\ 2018, \prl, 121, 062701

\bibitem[Lim \& Holt(2019)]{2019arXiv190205502L} Lim, Y., \& Holt, J.~W.\ 2019, arXiv:1902.05502

\bibitem[Lindblom(2010)]{2010PhRvD..82j3011L} Lindblom, L.\ 2010, \prd, 82, 103011

\bibitem[Lindblom \& Indik(2014)]{2014PhRvD..89f4003L} Lindblom, L., \& Indik, N.~M.\ 2014, \prd, 89, 064003

\bibitem[Ma et al.(2018)]{2018ApJ...858...74M} Ma, P.-X., Jiang, J.-L., Wang, H., et al.\ 2018, \apj, 858, 74

\bibitem[McNeil Forbes et al.(2019)]{2019arXiv190404233M} McNeil Forbes, M., Bose, S., Reddy, S., et al.\ 2019, arXiv:1904.04233

\bibitem[Most et al.(2018)]{2018PhRvL.120z1103M} Most, E.~R., Weih, L.~R., Rezzolla, L., \& Schaffner-Bielich, J.\ 2018, \prl, 120, 261103

\bibitem[N{\"a}ttil{\"a} et al.(2016)]{2016A&A...591A..25N} N{\"a}ttil{\"a}, J., Steiner, A.~W., Kajava, J.~J.~E., Suleimanov, V.~F., \& Poutanen, J.\ 2016, \aap, 591, A25

\bibitem[Oertel et al.(2017)]{2017RvMP...89a5007O} Oertel, M., Hempel, M., Kl{\"a}hn, T., \& Typel, S.\ 2017, RvMP, 89, 015007

\bibitem[{\"O}zel \& Psaltis(2009)]{2009PhRvD..80j3003O} {\"O}zel, F., \& Psaltis, D.\ 2009, \prd, 80, 103003

\bibitem[{\"O}zel \& Freire(2016)]{2016ARA&A..54..401O} {\"O}zel, F., \& Freire, P.\ 2016, \araa, 54, 401

\bibitem[{\"O}zel et al.(2016)]{2016ApJ...820...28O} {\"O}zel, F., Psaltis, D., G{\"u}ver, T., et al.\ 2016, \apj, 820, 28

\bibitem[Raithel et al.(2017)]{2017ApJ...844..156R} Raithel, C.~A., {\"O}zel, F., \& Psaltis, D.\ 2017, \apj, 844, 156

\bibitem[Radice \& Dai(2019)]{2019EPJA...55...50R} Radice, D., \& Dai, L.\ 2019, EPJA, 55, 50

\bibitem[Read et al.(2009a)]{2009PhRvD..79l4032R} Read, J.~S., Lackey, B.~D., Owen, B.~J., \& Friedman, J.~L.\ 2009, \prd, 79, 124032

\bibitem[Read et al.(2009b)]{2009PhRvD..79l4033R} Read, J.~S., Markakis, C., Shibata, M., et al.\ 2009, \prd, 79, 124033

\bibitem[Steiner et al.(2010)]{2010ApJ...722...33S} Steiner, A.~W., Lattimer, J.~M., \& Brown, E.~F.\ 2010, \apj, 722, 33

\bibitem[Steiner et al.(2013)]{2013ApJ...765L...5S} Steiner, A.~W., Lattimer, J.~M., \& Brown, E.~F.\ 2013, \apjl, 765, L5

\bibitem[Steiner et al.(2016)]{2016EPJA...52...18S} Steiner, A.~W., Lattimer, J.~M., \& Brown, E.~F.\ 2016, EPJA, 52, 18

\bibitem[Tews et al.(2017)]{2017ApJ...848..105T} Tews, I., Lattimer, J.~M., Ohnishi, A., \& Kolomeitsev, E.~E.\ 2017, \apj, 848, 105

\bibitem[The PyCBC Team(2018)]{2018ascl.soft05030T} The PyCBC Team\ 2018, PyCBC: Gravitational-wave data analysis toolkit, ascl:1805.030


\bibitem[Yagi \& Yunes(2013)]{2013Sci...341..365Y} Yagi, K., \& Yunes, N.\ 2013, Sci, 341, 365

\end{thebibliography}
\end{document}